\documentclass[12pt]{iopart}

\usepackage{bm}
\usepackage{epsfig}
\usepackage{citesort}
\usepackage{graphicx}
\eqnobysec

\begin{document}


\begin{flushright}
{\large \tt LAPTH-1317/09}
\end{flushright}

\title[Boltzmann code optimisation]{Optimising Boltzmann codes for the
 {\sc PLANCK} era}

\author{ Jan~Hamann$^1$, Amedeo~Balbi$^{2,3}$, Julien
Lesgourgues$^{1,4,5}$ and Claudia Quercellini$^2$} 

\address{$^1$~LAPTh (Laboratoire d'Annecy-le-Vieux de Physique Th\'eorique,
CNRS UMR5108 \& Universit\'e de Savoie), BP 110, F-74941
Annecy-le-Vieux Cedex, France\\ $^2$~Dipartimento di Fisica,
Universit\`a di Roma ``Tor Vergata'', via della Ricerca Scientifica 1,
I-00133 Roma, Italy\\ $^3$~INFN Sezione di Roma ``Tor Vergata'', via
della Ricerca Scientifica 1, I-00133 Roma, Italy \\ $^4$~\'Ecole
Polytechnique F\'ed\'erale de Lausanne, FSB/ITP/LPPC, BSP, CH-1015
Lausanne, Switzerland\\ $^5$~PH-TH, CERN, CH-1211 Gen\`eve 23,
Switzerland}

\ead{\mailto{jan.hamann@lapp.in2p3.fr},
     \mailto{amedeo.balbi@gmail.com},
     \mailto{julien.lesgourgues@lapp.in2p3.fr},
     \mailto{claudia.quercellini@gmail.com}}

\begin{abstract}
High precision measurements of the Cosmic Microwave Background (CMB)
anisotropies, as can be expected from the {\sc planck} satellite, will
require high-accuracy theoretical predictions as well.  One possible
source of theoretical uncertainty is the numerical error in the output
of the Boltzmann codes used to calculate angular power spectra.  In
this work, we carry out an extensive study of the numerical accuracy
of the public Boltzmann code \texttt{CAMB}, and identify a set of
parameters which determine the error of its output.  We show that at
the current default settings, the cosmological parameters extracted
from data of future experiments like Planck can be biased by several
tenths of a standard deviation for the six parameters of the standard
$\Lambda$CDM model, and potentially more seriously for extended
models.  We perform an optimisation procedure that leads the code to
achieve sufficient precision while at the same time keeping the
computation time within reasonable limits.  Our conclusion is that the
contribution of numerical errors to the theoretical uncertainty of
model predictions is well under control -- the main challenges for
more accurate calculations of CMB spectra will be of an astrophysical
nature instead.

\end{abstract}
\maketitle

\section{Introduction}\label{sec:introduction}

Any meaningful quantitative analysis of experimental data is based on
a comparison with the predictions of a theoretical model, and the
analysis of Cosmic Microwave Background (CMB) anisotropies is no
exception to this rule.  The amount of information to be gained from
observations is limited not only by experimental uncertainties (such
as detector noise, foregrounds, etc.), but also by the ability to
accurately predict observable quantities from a given theory.  There
are various sources of theoretical uncertainties.  Some, such as
cosmic variance, are endemic to the problem, and unavoidable.  Others
are based on insufficient theoretical understanding of the complex
processes involved~\cite{Hu:1995fqa} (examples include the physics of
recombination \cite{Fendt:2008uu}, reionisation~\cite{Lewis:2006ym},
and contributions due to the Sunyaev-Zel'dovich effects).
Additionally, the analysis may be compromised by inaccuracies in the
numerical programmes used to calculate the anisotropy angular power
spectra~\cite{Seljak:2003th}.

With increasingly sophisticated experiments, and the contribution from
experimental errors becoming less and less important, the relative
significance of theoretical errors increases. In fact, for the {\sc
planck} satellite \cite{planck}, the signal uncertainty of the
temperature anisotropies will be dominated by cosmic variance instead
of noise over a wide range of scales up to multipoles of $\ell \simeq
2000$.

Ignoring any type of uncertainty can lead to biased estimates of
parameter values, and, in the worst case, a wrong physical
interpretation of the data.  It is therefore imperative to either make
sure that the errors are small enough to be negligible, or, if that
should not be the case, to devise an appropriate strategy to deal with
the problem.

In the present work, we shall consider the numerical accuracy of the
Boltzmann codes employed to calculate the angular anisotropy spectra
$\mathcal{C}_\ell$ for given input values of cosmological parameters.
The first public Boltzmann code was released more than a decade
ago~\cite{Bertschinger:1995er}, and to date, there are several other
such programmes freely available for
download~\cite{Seljak:1996is,Lewis:1999bs,Doran:2003sy}.  The output
of all these codes is necessarily inaccurate to some extent, due to
the use of semi-analytical approximations as well as artifacts of the
numerical implementation, such as finite integration steps or the need
to interpolate.  These effects can be parameterised by a set of
\emph{accuracy parameters}, whose settings determine the accuracy of
the output, but also the computation time.  Here, we will focus our
analysis on the \texttt{CAMB} code by Lewis, Challinor and
Lasenby~\cite{Lewis:1999bs} in order to avoid possible systematic
effects caused by differences between codes -- a comparison of
different (more or less) independent Boltzmann codes was performed by
Seljak et al.~\cite{Seljak:2003th}, who found an excellent qualitative
agreement.\footnote{We verified that after updating various parts of
\texttt{CMBfast} (values of physical constants, recombination code,
etc.), the output of \texttt{CAMB} and \texttt{CMBfast} agree
sufficiently well.}  We extend their line of reasoning and present a
detailed analysis of the potential effects of numerical inaccuracies
on parameter estimates from {\sc planck} data.  Additionally, we
optimise the accuracy settings of \texttt{CAMB} to find an ideal
balance between precision and execution time.

We proceed in Section~\ref{sec:method} by defining an appropriate
measure of accuracy, identifying the relevant parameters which affect
the accuracy of the output spectra and constructing a set of CMB
\emph{reference spectra}. In Section~\ref{sec:optimisation}, we will
describe our optimisation procedure and present a recommended set of
accuracy parameters for \texttt{CAMB}, followed by an analysis of the
expected potential bias on the cosmological parameters of the vanilla
model caused by numerical inaccuracies in Section~\ref{sec:bias}.  The
impatient reader may prefer to skip directly to our conclusions in
Section~\ref{sec:conclusions}.

\section{Reference spectra}\label{sec:method}

\subsection{The fiducial model}

In principle, the impact of individual accuracy parameters and the
number of relevant parameters will depend on the underlying
cosmological model assumed, and on the values of the cosmological
parameters.  In this analysis, we will stick to the 6-parameter
$\Lambda$CDM-``vanilla''-model and we limit ourselves to a point in
the space of cosmological parameters that lies close to the best fit
to the WMAP 5-year data~\cite{Dunkley:2008ie}, see
Table~\ref{table:params}.  We neglect the effects of weak
gravitational lensing on the CMB spectra \cite{Lewis:2006fu} (which
will of course have to be taken into account in an analysis of real
{\sc planck} data), and ignore signatures of non-minimal models, like
massive neutrinos, tensor modes, spatial curvature, etc., and defer
their treatment to future work.

\begin{table}[t]
  \caption{Free cosmological parameters of the model, fiducial values
  used to generate the mock data, and prior ranges adopted in the
  analysis of Section~\ref{sec:bias}.\label{table:params}}\vskip5mm
  \hskip5mm
\footnotesize{\begin{tabular}{llll}
 \br
 Parameter&&Fiducial Value&Prior Range\\
 \mr
 Dark matter density & $\Omega_{\rm dm} h^2$ & 0.10976 & $0.01\to0.99$ \\
 Baryon density & $\Omega_{\rm b} h^2$ & 0.02303  & $0.005 \to 0.1$ \\
 Hubble parameter & $h$ & 0.7 & $0.4\to1$ \\
 Optical depth to reionisation & $\tau$  & 0.09 & $0.01 \to 0.5$ \\
 Scalar spectral index & $n_\mathrm{S}$ & 0.96
& $0.5\to 1.5$\\
 Amplitude of scalar spectrum @ $k = 0.05$ Mpc$^{-1}$& $\ln \left[
10^{10} A_\mathrm{S} \right]$
& 3.135 & $3\to 4$\\
 \br
\end{tabular}}
\end{table}

\subsection{Measuring accuracy\label{subsec:chi2}}

In order to quantify the accuracy of the Boltzmann code
output $\mathcal{C}_\ell^{\rm out}$, we require two things: a
reference point $\mathcal{C}_\ell^{\rm ref}$ to compare with, and a
measure of accuracy.

The reference spectra $\mathcal{C}_\ell^{\rm ref}$ would ideally be
the exact prediction of the theory.  In practice however, we have to
make do with getting close enough to these ideal spectra.  We will
return to this issue and describe the construction of
$\mathcal{C}_\ell^{\rm ref}$ in~\ref{subsec:refspec}.

To measure accuracy, one might be tempted at first glance to look at
the relative difference of the spectra $(\mathcal{C}_\ell^{\rm
out}-\mathcal{C}_\ell^{\rm ref})/\mathcal{C}_\ell^{\rm ref}$ for each
$\ell$.  However, this approach does not properly take into account
the fact that the accuracy requirements are dependent on $\ell$ (due
to cosmic variance and experimental errors).  Additionally, the
spectra are merely an intermediate step in the inference process.  In
the end, we are interested in possible biases on cosmological
parameters, not the accuracy of the spectra.  A more meaningful
measure of deviation from the reference spectra is the effective
$\chi^2$, which can be obtained by using the reference spectra to
generate a fiducial data set, taking into account the experimental
errors of an experiment (or the projected errors in case of future
experiments), and ``fitting'' $\mathcal{C}_\ell^{\rm out}$ to these
data.

More formally, the effective $\chi^2$ is related to the likelihood
$\mathcal{L}$ and defined by
\begin{equation}
  \label{eq:chi2}
  \chi^2 = - 2 \ln \mathcal{L} = \sum_\ell (2 \ell +1) \left[ \Tr
    \left(\mathbf{\tilde{C}_\ell}^{-1} \mathbf{\hat{C}_\ell}\right) +
    \ln \frac{\left| \mathbf{\tilde{C}_\ell} \right|}{\left|
        \mathbf{\hat{C}_\ell} \right|}-2 \right]. 
\end{equation}
Here, $\mathbf{\tilde{C}_\ell}$ is the theoretical covariance matrix, and
its entries are taken to be the sum of the signal and noise power
spectra:
\begin{equation}
  \mathbf{\tilde{C}}_\ell = \left\{ \mathcal{C}_\ell^{XX'} +
      \mathcal{N}_\ell^{XX'} \right\},
\end{equation}
where the index $X$ runs over temperature ($T$) and polarisation
($E$), and for a fiducial data set, we can take the data covariance
matrix $\mathbf{\hat{C}_\ell}$ to be equal to $\left.
  \mathbf{\tilde{C}_\ell} \right|_{\theta_0}$, i.e., the theoretical
covariance matrix evaluated for the fiducial values of the
cosmological parameters. 

We take the noise to be isotropic and Gaussian; the noise power
spectrum is related to the experimental parameters of
Table~\ref{table:hfilfispecs} through
\begin{equation}
  \mathcal{N}_\ell^{X X'} = \delta_{X X'} \; \theta_{\rm beam}^2 \,
  \Delta_X^2 \, \exp \left[ \ell \left( \ell +1 \right) \frac{\theta_{\rm
      beam}^2}{8 \ln 2} \right].
\end{equation}
For more details see Refs.~\cite{Perotto:2006rj,Hamann:2007sk}. It
should be noted that the normalisation of Eq.~(\ref{eq:chi2}) is
chosen such that the total $\chi^2$ is zero when the output spectra
exactly match the reference spectra used to construct the fiducial data.

We generate the fiducial data set of $TT$-, $EE$-, and $TE$-spectra up
to $\ell = 3000$ using the code of Perotto et
al.~\cite{Perotto:2006rj}.  For simplicity we ignore the effects of
incomplete sky coverage due to masking the galaxy and point sources,
as well as anisotropic noise \cite{Hamimeche:2008ai}. To evaluate the
accuracy of \texttt{CAMB} in view of the expected data from {\sc
planck}, we assume 14 months of integrated observations in the 70~GHz
channel of LFI and the 100 and 143~GHz channels of the HFI instrument;
their specifications are taken from the {\sc planck} blue
book~\cite{planck} and listed in Table~\ref{table:hfilfispecs}.

\subsubsection{Interpretation of the $\chi^2$ measure}

As can be seen in Eq.~\ref{eq:chi2}, $\chi^2$ is directly related to
the likelihood, which, along with a choice of prior probability
densities on all cosmological parameters, leads to the posterior
probability density, from which constraints on parameters are
eventually derived.  Assuming flat priors on the parameters and a
multivariate Gaussian likelihood function, for a given numerical error
$\chi^2$, the bias on any cosmological parameter cannot exceed
$\sqrt{\chi^2}$ standard deviations in the worst case (i.e., when the
error in the angular power spectra can be exactly offset by changing
one of the cosmological parameters).  On the other hand, if the error
had no degeneracy with any cosmological parameter, the effect of a
non-zero $\chi^2$ would be just a constant offset to the likelihood
function, which would not have any effect on parameter inference.  In
realistic cases, the expected bias would lie somewhere in between.
For the parameters of the vanilla model, we will provide an estimate
of the bias in Section~\ref{sec:bias}.

\begin{table}[t]
  \caption{\label{table:hfilfispecs}
    List of technical specifications for the 70~GHz channel of the LFI
    and the 100 and 143~GHz channels of the HFI instrument:
    $\theta_{\rm beam}$ denotes the beam width, $\Delta_{T,P}$
    are the sensitivities per pixel and $\nu$ is the centre frequency
    of the channels.}\vskip5mm \hskip45mm 
\footnotesize{
\begin{tabular}{cccc}
\br
 $\nu$/GHz & $\theta_{\rm beam}$ & $\Delta_T$/$\mu$K & $\Delta_P$/$\mu$K\\
\mr
  70 & 14.0'& 12.8& 18.3\\
 100 & 9.5' & 6.8 & 10.9\\
 143 & 7.1' & 6.0 & 11.4\\
\br
\end{tabular}}
\end{table}

\subsection{The accuracy parameters}

The numerical accuracy of a Boltzmann code's output depends on many
factors, ranging from the use of analytical approximations to the
sampling various intermediate quantities in the calculation.  These
sources of numerical error can be quantified in terms of
\emph{accuracy parameters}, e.g., the number of samples used for
interpolating a particular quantity.  An increase in accuracy will
generally be accompanied by a longer computation time and possibly
higher requirements on the available computer memory.

We use the June 2008 version of
\texttt{CAMB}\footnote{\texttt{http://www.camb.info}} as a starting
point of our analysis.  The unmodified version of \texttt{CAMB} comes
with a set of three continuously adjustable accuracy
parameters:\footnote{There are also a few logical switches that are
relevant here; in our analysis we kept them fixed to
\texttt{accurate\_polarization=T}, \texttt{accurate\_reionization=T}
and \texttt{do\_late\_rad\_truncation=T}.}
\begin{itemize}
	\item{\texttt{l\_sample\_boost}: determines for which values
	of $\ell$ the $\mathcal{C}_\ell$} are actually calculated (the
	rest are interpolated).

	\item{\texttt{l\_accuracy\_boost}: determines the multipole at
	which the Boltzmann hierarchy for photons, neutrinos, etc., is
	cut off.}

	\item{\texttt{accuracy\_boost}: affects the setting of various
	time steps, samplings, etc.}
\end{itemize}
The latter two parameters affect several settings at once, so in the
interest of optimising the performance of \texttt{CAMB}, we split them
up into their constituents and treat them separately.  Apart from the
settings governed by these three parameters, we identified a few other
quantities which can affect the accuracy of the results, and should be
taken into account when optimising the code.  Altogether, we consider
a set of 19 accuracy parameters in our analysis, listed in
Table~\ref{table:accpar}.

All parameters are defined in such a way that setting them a value of
1 reproduces the results of the unmodified version of \texttt{CAMB},
and larger values correspond to better accuracy.  The constituents of
the  old \texttt{l\_accuracy\_boost} and \texttt{accuracy\_boost}
parameters are taken to multiply the old \texttt{l\_accuracy\_boost} and
\texttt{accuracy\_boost} parameters (e.g., setting
$\texttt{lmaxg}=\texttt{lmaxnr}=2$ produces the same effect as setting
$\texttt{l\_accuracy\_boost}=2$).  We modified the routine that
determines for which values of $\ell$ the $\mathcal{C}_\ell$ are
calculated: our parameter \texttt{new\_l\_sample\_boost} is defined to
be the square root of the old \texttt{l\_sample\_boost};  for
$\texttt{new\_l\_sample\_boost} > 5$, all $\mathcal{C}_\ell$ are
calculated and there will be no interpolation of the final spectrum.

\begin{table}[t]
  \caption{Individual accuracy parameters for \texttt{CAMB}-\texttt{CAMB} comparison.\label{table:accpar}}\vskip5mm \hskip-15mm
  \footnotesize{\begin{tabular}{lcl} \br
      Parameter& Corresponding old accuracy parameter & Description and comments \\
      \mr
      \texttt{new\_l\_sample\_boost}&  \texttt{l\_sample\_boost} & $\ell$-sampling of $\mathcal{C}_\ell$s\\
      \mr
      \texttt{lmaxg}& \texttt{l\_accuracy\_boost}& Boltzmann hierarchy cutoff for photons\\
      \texttt{lmaxnr}&& Boltzmann hierarchy cutoff for massless neutrinos\\
      \mr
      \texttt{int\_tol}& & tolerance parameter for integration routines\\
      \texttt{ri\_timestep}& & time step during reionisation\\
      \texttt{rec\_timestep}& & time step during recombination\\
      \texttt{rec\_timestep2}& & time step between recombination and reionisation\\
      \texttt{rad\_trunc}& & truncation of photon hierarchy during matter domination\\
      \texttt{dec\_start}&  \texttt{accuracy\_boost}& starting time of decoupling\\
      \texttt{int\_sample}&& samples for integration over source function\\
      \texttt{source\_dk}& & $k$-sampling of source function\\
      \texttt{source\_kmin}& & minimum value of $k$ to calculate source function for\\
      \texttt{int\_xlmax1}& &starting time for source function integration\\
      \texttt{tc\_largek}& & switch off tight coupling later for large $k$\\ \mr
      \texttt{tc\_ep0}& -- & tight coupling switch\\
      \mr
      \texttt{bess\_sampling}& -- &  $x$-sampling of spherical Bessel functions $j_\ell(x)$ \\
      \texttt{bess\_xlimmin}& -- & approximate $j_\ell(x) \simeq 0$ for small $x$, if $\ell \geq \texttt{xlimmin}$\\
      \texttt{bess\_xlimfrac} & --  &  approximate $j_\ell(x)\simeq 0$ for large $\ell$, if $x \leq (1- \texttt{xlimfrac}) \cdot \ell$\\
      \mr
      \texttt{ketamax}& -- & maximum value of $k \eta$\\ \br
\end{tabular}}
\end{table}

\subsection{Constructing a reference data set \label{subsec:refspec}}

To quantify absolute accuracy we require a reference data set, as
discussed above.  Its construction is naturally tied to choosing the
accuracy parameters in such a way that increasing them further would
not have any appreciable effect.  However, by arbitrarily increasing
all parameters to ``large'' values at the same time, one would run
into the limits of the hardware, particularly the available memory.
For the purpose of finding suitable values for generating the
reference spectra, we therefore analyse the parameters one by one,
keeping all other parameters fixed.  For each parameter, we generate a
fiducial reference data set with that parameter set to a high value,
all other parameters kept at a value of 2.  Varying this parameter and
calculating the $\chi^2$ reveals its impact on overall accuracy,
allows us to find a suitable setting for the reference spectra and
lets us estimate the remaining error.

\begin{figure}[h!]
  \caption{\label{fig:scan1} These diagrams illustrate the dependence
  of accuracy on the settings of individual parameters.  We plot
  $\chi^2$ as a function of parameter value.  The fiducial reference
  data sets were generated with all parameters set to a value of 2,
  except for the one scanned, which is set to an extremely high value
  (100 for \texttt{dec\_start}, 5 for \texttt{tc\_largek}, and 10 for
  all other parameters).  When setting the accuracy parameters to
  their fiducial values, we obtain a $\chi^2$ of order $10^{-8}$
  instead of 0.  This effect stems from a numerical rounding error
  when outputting the fiducial data set, and is small enough not to be
  of relevance to any of our conclusions.}
\begin{center}
\includegraphics[width=\textwidth]{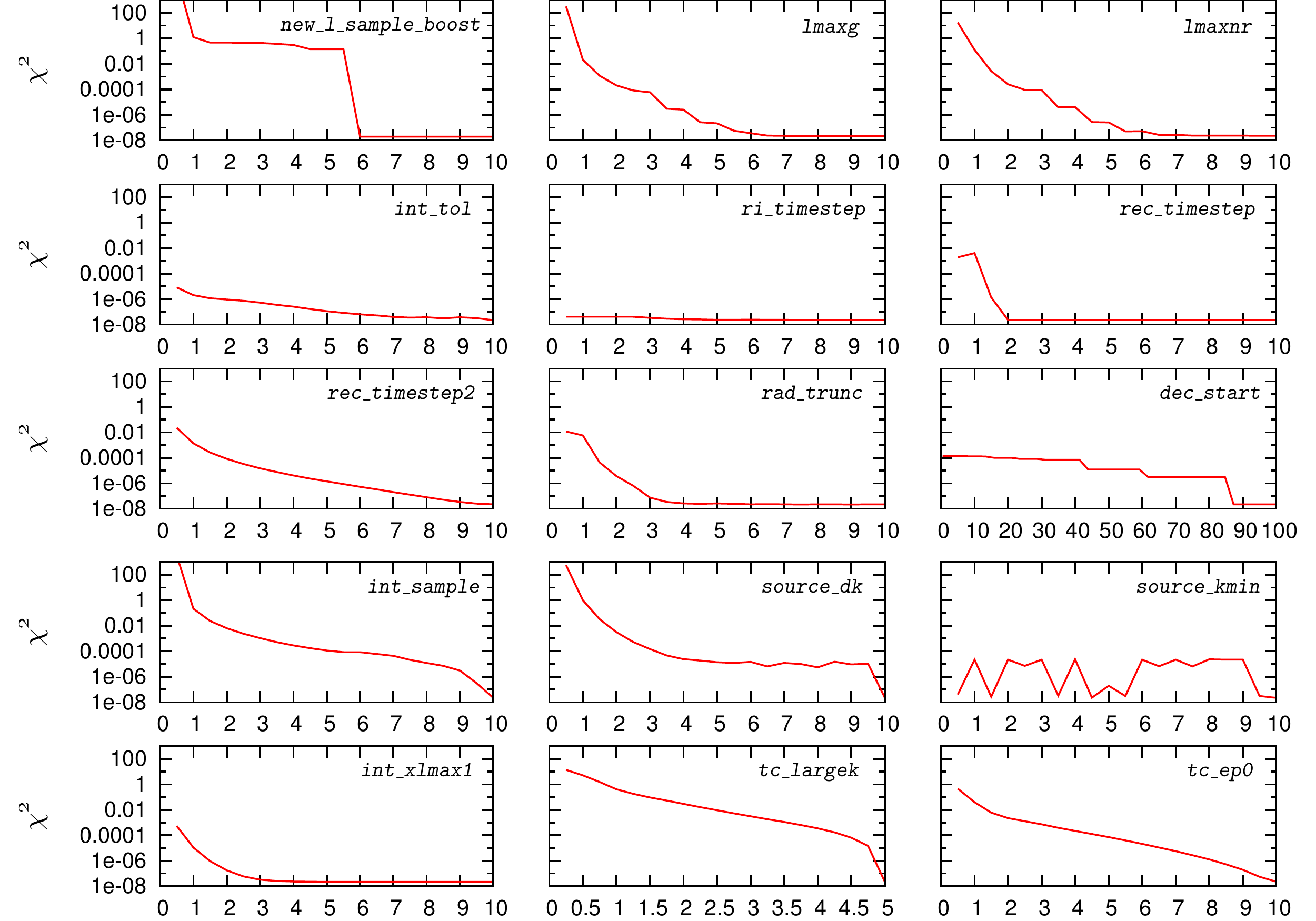}
\includegraphics[width=\textwidth]{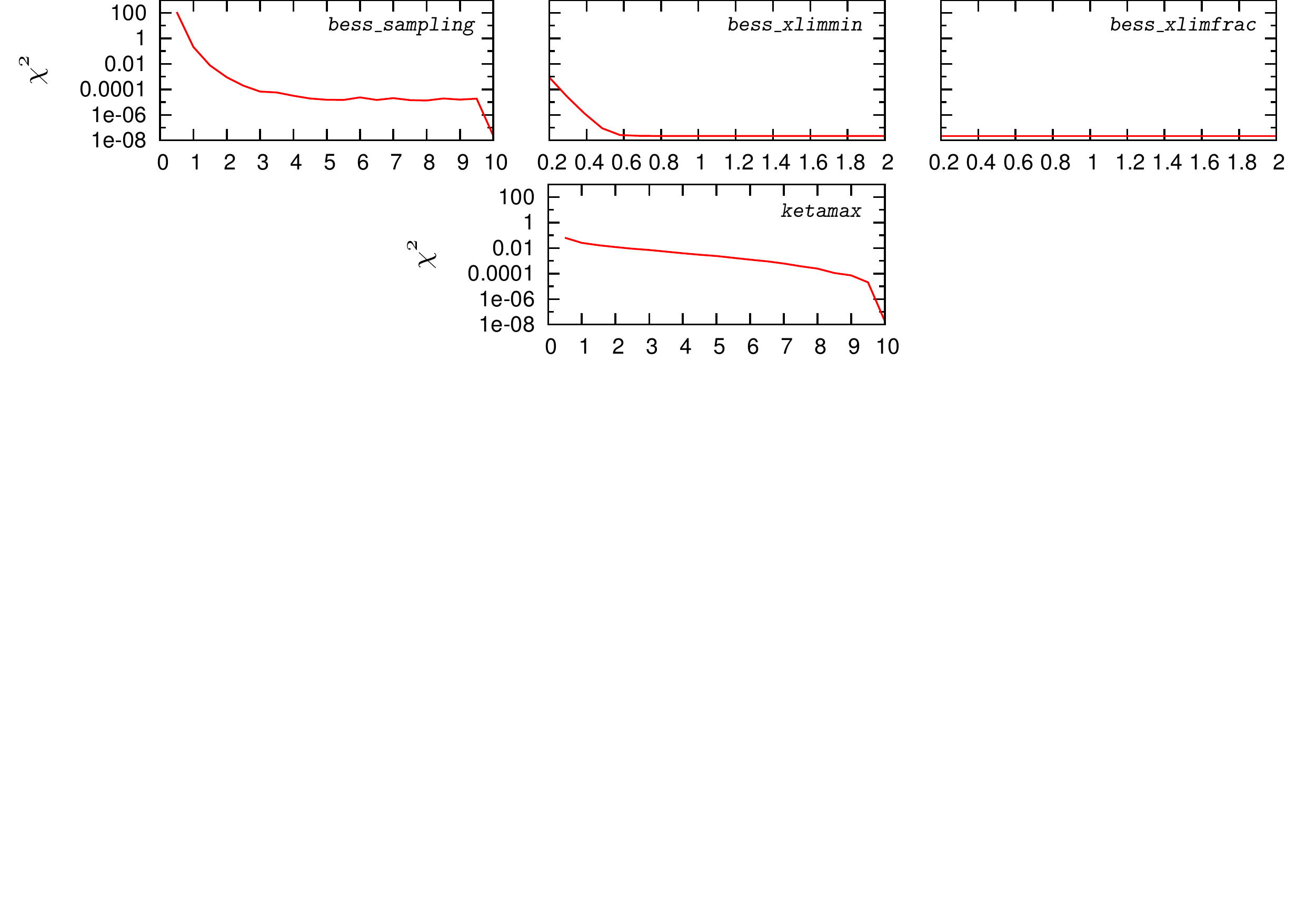}
\end{center}
\end{figure}

\begin{figure}[h!]
  \caption{\label{fig:scan2} These diagrams show the computation time
  $T$ (in seconds), varying the settings of all parameters
  individually while keeping all others fixed at a value of 2.  The
  calculations were performed on a single core of a $2.40\rm{\ GHz}$
  Intel T7700 CPU.  The spikes for \texttt{new\_l\_sample\_boost}, as
  well as the ``bumps'' at low settings of various other parameters
  are due to the fact that the Bessel functions were (re-)calculated
  at these points, adding a few seconds to the total time.}
\begin{center}
\includegraphics[width=\textwidth]{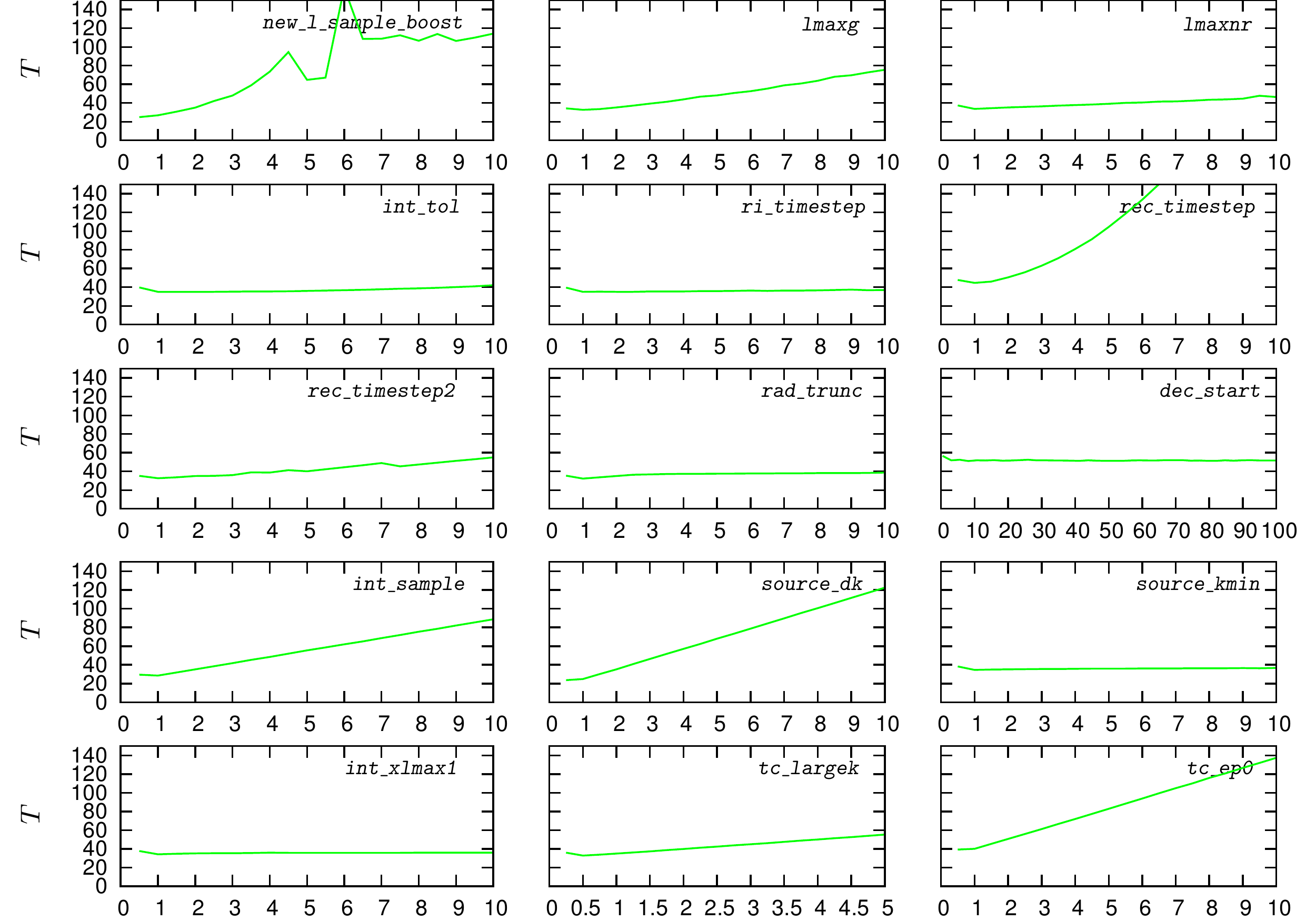}
\includegraphics[width=\textwidth]{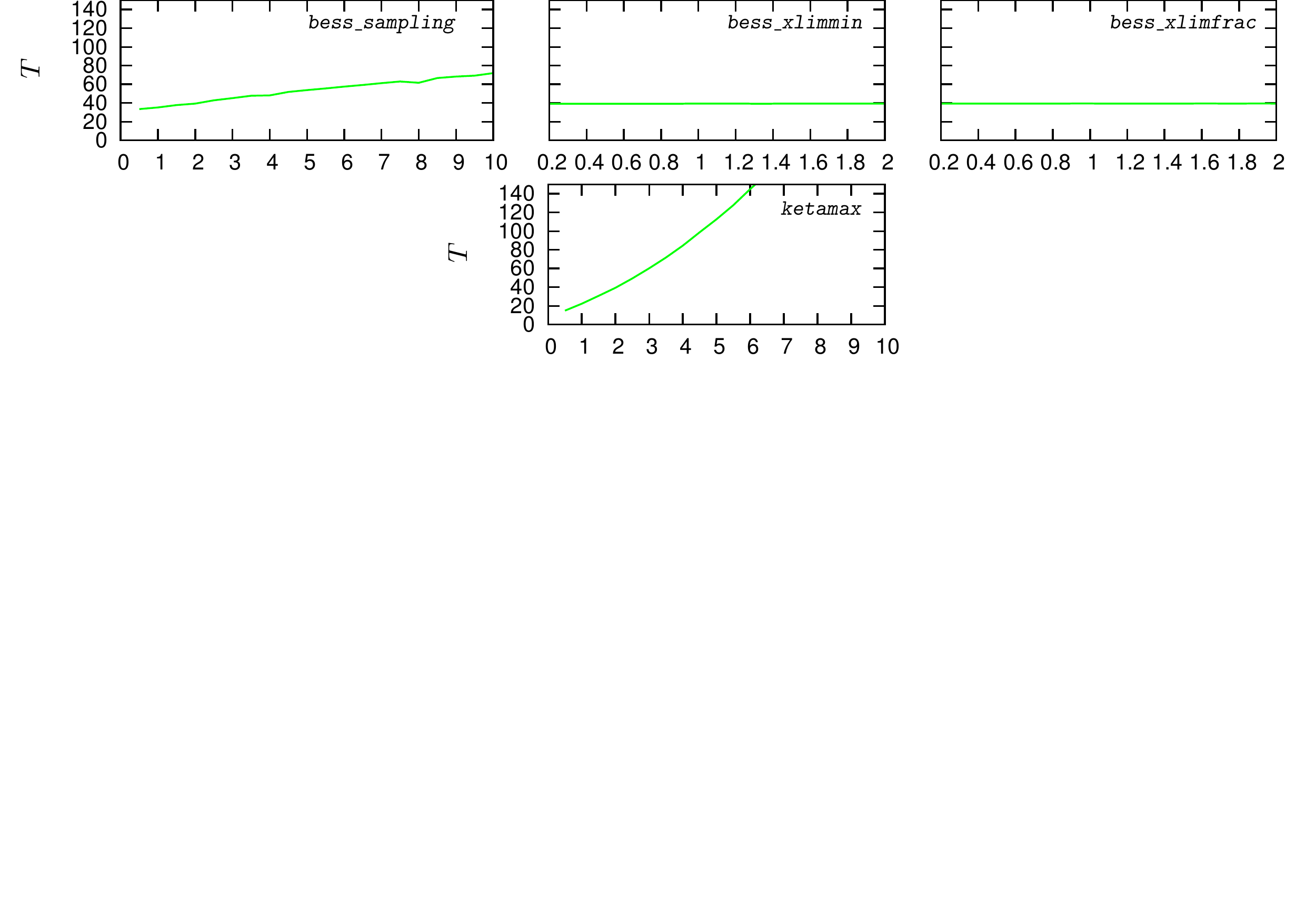}
\end{center}
\end{figure}

We show the results of the single parameter scans in
Figure~\ref{fig:scan1}-\ref{fig:scan2}. From Figure~\ref{fig:scan1} we
can see that certain parameters (e.g., \texttt{lmaxg},
\texttt{lmaxnr}) are very well-behaved and reach a
$\chi^2 \sim 10^{-8}$ plateau well below the fiducial value.  For
these parameters it is reasonable to assume that picking even higher
values would not have any appreciable effect on accuracy.  A number of
other parameters do not fully converge, and exhibit a step-like
behaviour before reaching the fiducial value (e.g.,
\texttt{int\_sample}, \texttt{source\_dk}).  It is likely that
increasing their value beyond our fiducial maximum value would have an
effect on the spectra.  However, the graphs in Figure~\ref{fig:scan1}
nonetheless provide an order of magnitude estimate of the remaining
error and one can use them as a guide to finding reasonable settings
for the construction of the reference spectrum.  Finally, the
parameter \texttt{source\_kmin} does not seem to converge at all and
exhibits unstable behaviour, though its effect on overall accuracy is
negligible.

The parameter settings we chose for the reference spectrum are given
in Table~\ref{table:settings}.  The dominant contribution to any
residual error of the reference spectrum will come from the parameter
displaying the worst convergence -- \texttt{ketamax}.  Unfortunately,
this parameter also has a strong impact on the computation time $T$
(see Figure~\ref{fig:scan2}), and memory requirements, precluding us
from choosing a higher setting.  We estimate the reference spectrum to
be accurate to $\Delta \chi^2$ of order $10^{-2}$.

\begin{table}[t]
  \caption{Accuracy settings for the reference spectrum (plus three recommended example
  settings, see Section~\ref{sec:optimisation}). \label{table:settings}}\vskip5mm
  \hskip-15mm \footnotesize{
\hspace{40mm}
\begin{tabular}{lcccc} \br
      Parameter& Reference setting & Setting 1 & Setting 2 & Setting 3 \\
      \mr
      \texttt{new\_l\_sample\_boost}& 6 & 1.77 & 1.60 & 2.00\\
      \mr
      \texttt{lmaxg}& 6 & 7.39 & 2.02 & 2.05\\
      \texttt{lmaxnr}& 6 & 1.78 & 2.11 & 5.36\\
      \mr
      \texttt{int\_tol}& 6 & 2.07 & 1.53 & 5.86\\
      \texttt{ri\_timestep}& 2 & 0.99 & 0.71 & 2.87\\
      \texttt{rec\_timestep}& 3 & 0.50 & 0.50 & 0.75\\
      \texttt{rec\_timestep2}& 10 & 1.85 & 1.14 & 2.44\\
      \texttt{rad\_trunc}& 4 & 1.81 & 1.76 & 2.58\\
      \texttt{dec\_start}& 100 & 2.35 & 35.76 & 5.05\\
      \texttt{int\_sample}& 6 & 4.21 & 4.88 & 3.65\\
      \texttt{source\_dk}& 4 & 3.10 & 2.84 & 2.72\\
      \texttt{source\_kmin}& 1 & 5.86 & 2.50 & 5.23\\
      \texttt{int\_xlmax1}& 4 & 1.20 & 1.00 & 2.16\\
      \texttt{tc\_largek}& 5 & 2.44 & 2.33 & 1.90\\
      \mr
      \texttt{tc\_ep0}& 5 & 4.25 & 3.30 & 6.32\\ 
      \mr
      \texttt{bess\_sampling}& 3 & 3.02 & 2.53 & 4.00 \\
      \texttt{bess\_xlimmin}& 2 & 1.14 & 0.74 & 3.61\\
      \texttt{bess\_xlimfrac} & 2 & 4.19 & 0.90 & 1.07\\
      \mr
      \texttt{ketamax}& 3 & 1.32 & 0.96 & 0.56\\ \br
\end{tabular}}
\end{table}

\section{Optimising performance}\label{sec:optimisation}

\begin{figure}[h]
  \caption{\label{fig:scatter} Computation time versus $\chi^2$ for
  the $\sim 20000$ samples of our Monte Carlo run.}
\begin{center}
\includegraphics[width=0.8\textwidth]{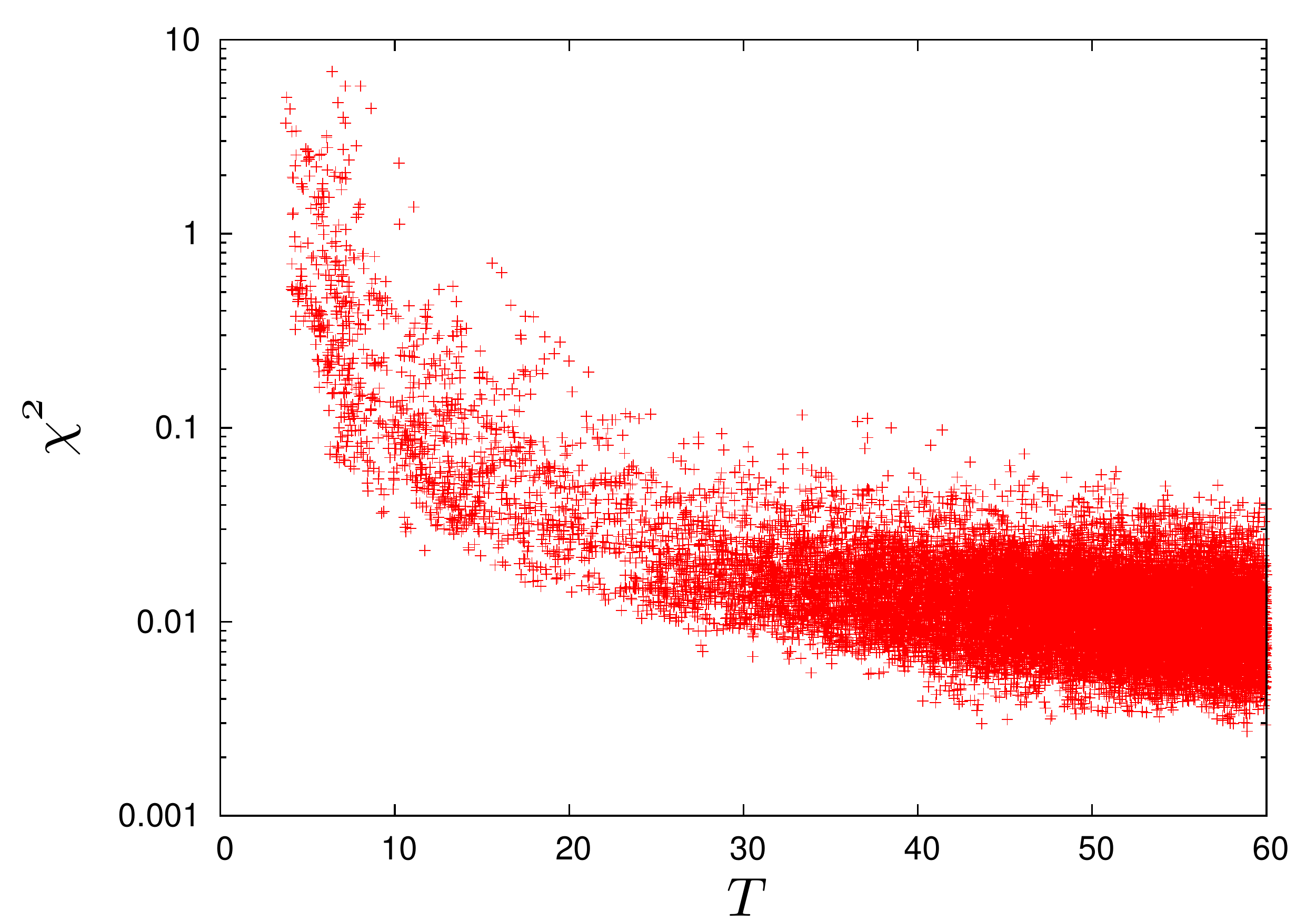}
\end{center}
\end{figure}

Having constructed the reference spectra, we can now proceed to
searching settings of the accuracy parameters which yield a result
that lies as close as possible to the reference spectra, within a
reasonable time of computation.  To this end, we use a modified
version of the Markov chain Monte Carlo code
\texttt{CosmoMC}~\cite{Lewis:2002ah}.  Major modifications include:
\begin{itemize}
	\item{we use the fiducial data set constructed from the
	reference spectra;}

	 \item{we vary the accuracy parameters instead of the
	 cosmological parameters (which are kept fixed at their
	 fiducial values), taking top hat priors on all accuracy
	 parameters (with lower limits at a value of 0.5 and upper
	 limits large enough to not influence the results);}

	\item{instead of sampling from the usual posterior
	$\mathcal{P}$ (which is proportional to \mbox{$\mathcal{L}
	\sim \exp[-\chi^2/2]$}) itself, we sample a function
	$F(\mathcal{P})$, defined by 

	\begin{equation} 
	 F(\mathcal{P}) = \left\{ \begin{array}{c c l}
	 0 & \quad \mbox{if} \; & \mbox{$T > 60\ \rm{s}$}\\
	 \mathcal{P} & \quad \mbox{if} \; & \mbox{$\mathcal{P} \geq 1$}\\
	 5 \ln {\left[\mathcal{P}\right]} + 1 & \quad \mbox{if} \; &
	 \mbox{$\mathcal{P} < 1$}\\ \end{array} \right.
	\end{equation}

	 This function was chosen such that areas of parameter space
	 leading to too long computation times are avoided, and that
	 areas of parameter space giving $\chi^2 \ll 1$ are better
	 sampled.}
\end{itemize}
We generated $\sim \! \! 20000$ sample settings in this way; a scatter
plot of the samples in the $(\chi^2, T)$-plane is presented in
Figure~\ref{fig:scatter}, illustrating the strong correlation between
accuracy and computation time.  Which settings to use is a somewhat
subjective decision and should be taken with one's available computing
power in mind.  We list two sample settings in
Table~\ref{table:settings}, taken from the samples of our Monte Carlo
run: ``Setting 1'' corresponds to the best accuracy under the
constraint that the computing time be less than 60 seconds, ``Setting
2'' gives the best accuracy for $T < 30\ {\rm s}$, and ``Setting 3''
the best accuracy for $T < 17.6\ {\rm s}$ (which is the computation
time for the unmodified version of \texttt{CAMB} run at
\texttt{accuracy\_level = 2}).  The performance of
these settings is contrasted to \texttt{CAMB}'s default
(\texttt{accuracy\_level = 1}), and a high-accuracy setting of an
unmodified version of \texttt{CAMB} (\texttt{accuracy\_level = 2}) in
Table~\ref{table:improvement}. 

Note that at the default settings, with a difference of $\Delta
\chi^2$ to the reference spectra, parameter estimates could be biased
by more than two standard deviations, in the worst case.  For both
Settings 1 and 2, on the other hand, the maximum possible bias would
be less than 0.1 standard deviations ($\sim 0.13$ standard deviations
for Setting 3), assuming Gaussian posterior distributions.

The results of the MCMC search confirm the tendencies observed in the
single parameter scans of Figs.~\ref{fig:scan1} and~\ref{fig:scan2}
regarding the impact of individual parameters on accuracy and speed;
we find no evidence for significant cross-correlations between
accuracy parameters.

\begin{table}[t]
  \caption{Deviation from reference spectrum, given in terms of
  $\chi^2$, and computation time $T$, for the three example settings of
  Table~\ref{table:settings}, \texttt{CAMB}'s default settings
  ($\texttt{accuracy\_level} = 1$) and $\texttt{accuracy\_level} =
  2$. \label{table:improvement}}\vskip5mm \hskip-15mm \footnotesize{
\hspace{60mm}
\begin{tabular}{lcc} \br
      Setting & $\chi^2$ & $T\cdot s^{-1}$ \\
      \mr
	1 & $2.7 \cdot 10^{-3}$  & 59 \\
	2 & $7.0 \cdot 10^{-3}$ & 27 \\
	3 & $1.6 \cdot 10^{-2}$ & 17.4 \\
	$\texttt{accuracy\_level} = 1$ & 5.8 & 2.7 \\
	$\texttt{accuracy\_level} = 2$ & 0.5 & 17.6 \\ \br
\end{tabular}}
\end{table}

\section{Estimating the bias}\label{sec:bias}

\begin{figure}[h]
  \caption{\label{fig:bias} Marginalised posterior probability
  densities of the cosmological parameters of the vanilla model.
  Thick black lines correspond to the exact results (using the same
  accuracy settings for the MCMC and for the fiducial data set), green
  lines are results obtained with the ``Settings 2'' column of
  Table~\ref{table:settings} and the reference data set, and the red
  dotted lines represent the posteriors from the default accuracy
  settings of \texttt{CosmoMC}/\texttt{CAMB} and the reference data set.}
\begin{center}
\includegraphics[width=\textwidth]{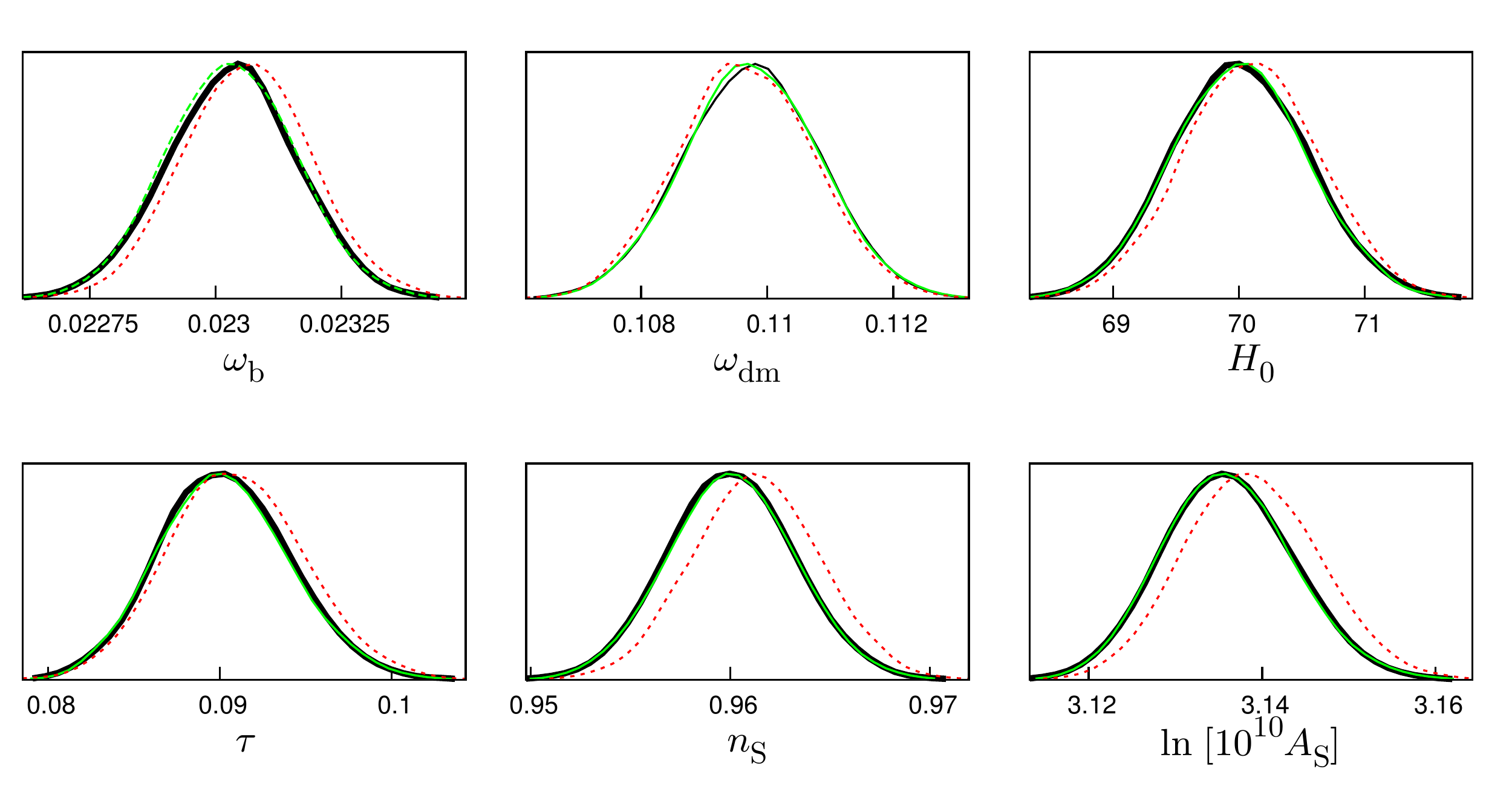}
\end{center}
\end{figure}

As mentioned above, the $\chi^2$ for a given accuracy settings can be
used to estimate the worst case bias on cosmological parameters.
Nevertheless, we are also interested in how large a bias effect we can
expect in the example case of the vanilla model.  To this end, we
performed an actual parameter estimation exercise using
\texttt{CosmoMC}, for three cases:
\begin{itemize}
	\item{Using the same \texttt{CAMB} accuracy settings for
	generating the fiducial data as for the inference process --
	mimicking an analysis free of numerical errors.}  

	\item{Using the reference data set and running \texttt{CAMB}
	at accuracy settings ``2'' from Table~\ref{table:settings}.}
	
	\item{Using the reference data set and running \texttt{CAMB} at
	its default settings.}
\end{itemize}
We generate 16 Markov chains, making sure the Gelman-Rubin convergence
parameter $R-1$ \cite{gelru} is smaller than $10^{-2}$ for all
parameters considered.  The results are plotted in
Figure~\ref{fig:bias}: as expected, there is no discernible bias
between the results obtained with optimised settings and an error-free
analysis.  The expected bias for the default settings is rather mild,
not exceeding a few tenths of the posterior standard deviations for
the respective parameters, with $n_{\rm S}$ (40\%) and $\ln
\left[10^{10} A_{\rm S}\right]$ (43\%) being the most affected.

\section{Conclusions}\label{sec:conclusions}

To meet the challenge posed by ultra-precise future CMB experiments,
the output of the Boltzmann codes used to calculate theoretical
predictions for the CMB anisotropy spectra will have to meet stringent
requirements in terms of numerical accuracy -- a goal that comes at
the cost of accordingly higher demand in computational resources.
This tendency is partially mitigated by the increase in available
computing power.  Additionally, the use of efficient interpolation
algorithms
\cite{Sandvik:2003ii,Fendt:2006uh,Auld:2006pm,Fendt:2007uu} can
significantly accelerate the process of parameter estimation from
future data sets.  Nonetheless, such methods still rely on the input
of a full Boltzmann code, whose inherent numerical errors will be
propagated to the interpolation codes.

We have performed a detailed analysis of the numerical accuracy of
\texttt{CAMB}, provided an estimate of the residual numerical error of
the output, and evaluated the possible impact on parameter estimates
from {\sc planck} data.  For the present default accuracy settings of
\texttt{CAMB} and simulated {\sc planck} data, numerical errors can
lead to a slight bias on estimates of the six free parameters of the
$\Lambda$CDM model; though we cannot exclude the possibility of a more
serious bias on parameters of extended models.  However, by tweaking
the settings of various internal parameters of \texttt{CAMB} it can be
made sure that the bias on any parameter will not exceed 0.1 standard
deviations.
An example for suggested settings of the
accuracy parameters is given in Table~\ref{table:settings}.

The results of this paper lead to an efficient and more accurate
calculation of CMB angular power spectra, and should bring
\texttt{CAMB} to a standard that will allow us to make the most out of
upcoming {\sc planck} data.  We conclude that the contribution of
numerical errors to the theoretical uncertainty of model predictions
is well under control -- the main challenges for more accurate
calculations of CMB spectra will be of an astrophysical nature instead.

\section*{Acknowledgments}
We thank Martin Bucher, Anthony Challinor, Loris Colombo, Fabio
Finelli and Antony Lewis for discussions and helpful suggestions.

\section*{References}

\end{document}